\documentclass{optica-article}

\journal{opticajournal} 

\articletype{Research Article}

\usepackage{lineno}

\begin{document}

\title{Ultrafast Field-Resolved Nonlinear Optical Spectroscopy in the Molecular Frame}

\author{Siddhant Pandey,\authormark{1} Liang Z. Tan,\authormark{2} Francis Walz,\authormark{1} Varun Makhija,\authormark{3} and Niranjan Shivaram\authormark{1,4,*}}

\address{\authormark{1}Department of Physics and Astronomy, Purdue University, West Lafayette, IN 47907 USA\\
\authormark{2}Molecular Foundry, Lawrence Berkeley National Laboratory, Berkeley, California 94720 USA\\
\authormark{3}Department of Physics, University of Mary Washington, Fredericksburg, VA 2240 USA\\
\authormark{4}Purdue Quantum Science and Engineering Institute, Purdue University, West Lafayette, IN 47907 USA}

\email{\authormark{*}niranjan@purdue.edu} 


\begin{abstract*}
We resolve the real-time electric field of a femtosecond third-order nonlinear optical signal in the molecular frame. The electric field emitted by the induced third-order polarization from impulsively pre-aligned gas-phase molecules at room temperature, in a degenerate four-wave mixing (DFWM) scheme, is measured using a spectral interferometry technique. We show that by measuring both the amplitude and phase of the emitted femtosecond pulse, information related to electronic symmetries can be accessed. The nonlinear signal is measured around a rotational revival to extract its molecular-frame angle dependence from pump-probe time delay scans. By comparing these measurements for two linear molecules, carbon dioxide (CO\textsubscript{2}) and Nitrogen (N\textsubscript{2}), we show that the measured second-order phase parameter (temporal chirp) of the signal is sensitive to the valence electronic symmetry of the molecules, whereas the amplitude of the signal does not show such sensitivity. We compare these measurements to theoretical calculations of the chirp observable in the molecular frame. This work is an important step towards using field-resolved nonlinear optical measurements to study ultrafast dynamics in electronically excited molecules.
\end{abstract*}

\section{Introduction}
Ultrafast dynamics in molecules occur on time scales ranging from attoseconds to picoseconds. These dynamics are routinely studied using photoionization based spectroscopies, ultrafast electron diffraction, and all-optical spectroscopies \cite{krausz2009,geneaux2019,li2020,nisoli2017}. Due to the multidimensional nature of the problem, the study of ultrafast dynamics in molecules typically requires a number of complimentary measurements to disentangle the dynamics for any given system. An all-optical experimental observable that is sensitive to electronic symmetry could offer important insight into ultrafast electron and electron-nuclear dynamics. Ultrafast optical measurements, including transient absorption spectroscopy, rely on the nonlinear optical response of the target molecule. In symmetric top molecules, the two unique components of the polarizability tensor, $\alpha_{\parallel}$ and $\alpha_{\perp}$, contain limited information compared to the multiple nonzero tensor components of higher-order hyperpolarizability tensors. A measurement that probes higher-order nonlinear response of molecules can thus provide detailed information on the symmetry of involved electronic states. Further, in an all-optical measurement, completely resolving the emitted electric field (E-field) provides direct access to the induced polarization which is intricately related to the ultrafast evolution of the system being studied. Combining ultrafast field-resolved spectroscopy with nonlinear optical response measurements will thus enable tracking of transient electronic symmetries in excited molecules. Recently, field-resolved ultrafast measurements using attosecond streaking \cite{itatani2002,goulielmakis2004,eckle2008}, direct field sampling \cite{liu2021,liu2022,bionta2021,park2018,hui2022,schiffrin2013,paasch2014,paasch2016,sederberg2020} and spectral interferometry \cite{walz2022} have emerged as sensitive methods to probe ultrafast dynamics. Applying field-resolved nonlinear optical spectroscopy to laser excited molecules is an important step towards realizing the full potential of nonlinear optical spectroscopy in probing ultrafast dynamics.

Due to inversion symmetry, typically, the first non-vanishing nonlinear response in gas-phase molecules is the third-order response, corresponding to molecular second-hyperpolarizabilities. In this work, we measure the molecular frame third-order response in gas-phase linear molecules at room temperature, by directly measuring the full nonlinear E-field emitted during degenerate four-wave mixing (DFWM) \cite{shirley1980} in pre-aligned molecules. This nonlinearity has three dominant sources - bound electronic, plasma, and rotational nonlinearity \cite{wahlstrand2013}. Vibrational nonlinearities are not observed due to the limited bandwidth of our laser pulses. Rotational nonlinearity, which arises from nuclear motion, is slower compared to the near-instantaneous electronic nonlinearity, which arises from the distortion of the molecular electron cloud due to the laser's electric field \cite{chen2007}. For low enough laser intensities, the plasma nonlinearity can be ignored. When all the DFWM laser pulses have polarization along a fixed axis in the lab frame, say $z$, the emitted third-order signal in the frequency domain from a single molecule can be written as

\begin{equation}\label{eq:esignal}
    E_{signal,z}(\omega,\theta) = i \chi^{(3)}_{zzzz}(\omega, \theta) E_{1,z}(\omega) E_{2,z}^*(\omega) E_{3,z}(\omega)
\end{equation}

where $\theta$ is the relative angle between the laser polarization along $\hat{z}$ and the molecules' symmetry-axis, $\omega$ is the angular frequency, the subscripts 1, 2 and 3 correspond to the three DFWM pulses which are assumed to be temporally overlapped with zero time delay. For linear molecules, which will be the focus of this work, the lab frame third-order susceptibility can be related to the molecular frame second-hyperpolarizabilities as

\begin{equation}\label{eq:chi3}
    \chi^{(3)}_{zzzz}(\omega, \theta) = \gamma^{(2)}_{zzzz}(\omega) \cos^4(\theta) + \frac 3 2 \gamma^{(2)}_{zzxx}(\omega) \sin^2(2\theta) + \gamma^{(2)}_{xxxx}(\omega) \sin^4(\theta)
\end{equation}

It is well known that molecules can be excited rotationally with intense non-resonant laser pulses, leading to periodic rotational revivals on the time scale of tens of picoseconds \cite{stapelfeldt2003,rosca2001,renard2004,renard2003}. Electron dynamics, on the other hand, occur on femtosecond and attosecond time scales after interaction with the excitation laser pulse. This separation of time scales allows probing of femtosecond electronic response using DFWM at rotational revivals by first exciting a rotational wavepacket. Once the DFWM input pulses are characterized, a measurement of $E_{signal}(t)$ from a rotational wavepacket can give direct access to molecular frame second-hyperpolarizability. This is essentially similar to measuring the lab-frame nonlinear response in equation \ref{eq:chi3} for multiple $\theta$ to get the molecular frame response tensor components.

\section{\label{}Experimental Method}
\begin{figure}[t]
\centering
\includegraphics[width= 1.0 \textwidth]{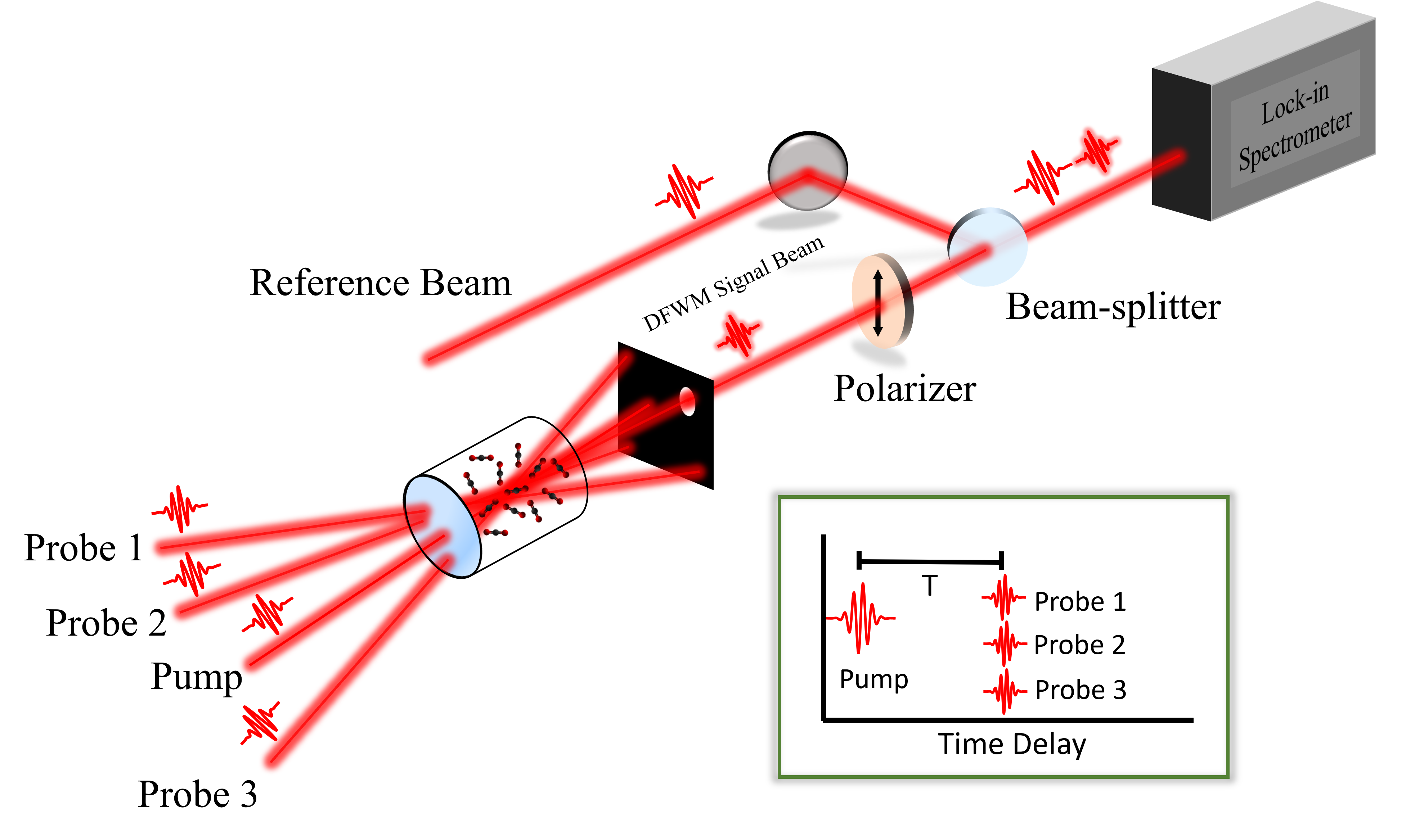}
\caption{Schematic of the experimental setup. The alignment Pump and the time-delayed DFWM Probe beams are focused into a gas cell containing the target gas at room temperature and a pressure of 4 bar. The emitted nonlinear signal is spatially isolated, cleaned with a polarizer, and combined with the external reference in a lock-in detection enabled spectrometer. The Reference is separately characterized using a Frequency Resolved Optical Gating (FROG) setup (not shown).}
\label{fig1:setup}
\end{figure}

In our field-resolved alignment pump-DFWM experiment (see figure \ref{fig1:setup}), 60 fs near-infrared (IR) pulses centered around 800 nm are first split and delayed. One arm forms the alignment pump beam and the other is split again into three weaker DFWM probe beams using a mask, in the folded BOXCARS geometry \cite{shirley1980}. One of the DFWM probe beams is further split to derive a reference pulse. The alignment pump excites a rotational wavepacket which is then probed using the DFWM beams. All four pulses intersect inside a gas cell containing the target gas at room temperature and a pressure of 4 bar, in a non-collinear geometry. The intensity of the pump pulse was estimated from the fitting procedure to be 8 TW/cm$^2$, while the average intensity of the probe pulses is estimated to be $<4$ TW/cm$^2$. The crossing angles are small enough such that time-smearing is small in comparison to the pulse duration. Time delay (T) between the alignment pump and probe pulses is varied using an optical delay stage. The relative polarization of the pump and probe beams is set to $0^\circ$. In the folded BOXCARS geometry, the emitted nonlinear signal propagates along a separate direction, and is spatially isolated from all other beams using a beam-stop \cite{shirley1980}. The emitted signal is passed through a polarizer to remove any ellipticity and coupled into a home-built spectrometer, along with the reference pulse, for spectral interferometry \cite{fittinghoff1996}. In our measurements, the DFWM signal from aligned molecules is $\sim 1\%$ of the signal from unaligned molecules. Since both these travel along the same phase-matched direction, it becomes essential to remove this background signal from unaligned molecules. We adapt a lock-in amplification scheme in our spectrometer to separate the weak signal from the strong background. The details of this lock-in enabled interferometry scheme will be discussed in a future publication. The use of a lock-in spectrometer results in a significant improvement of the signal-to-noise ratio (SNR), and automatic subtraction of the background nonlinear signal from unaligned molecules.


For each pump-probe time delay, the measured E-field phase is fit to a $5^{th}$ order polynomial

\begin{equation}\label{eqn:phase}
    \varphi(t,T) = a_0(T) + a_1(T) \cdot t + a_2(T) \cdot t^2 + a_3(T) \cdot t^3 ...
\end{equation}

similar to reference \cite{walz2022}. The second-order polynomial coefficient $a_2$ (also known as chirp), which is the dominant nonlinear fit coefficient, is extracted as a function of the pump-probe time delay ($T$). Experiments that measure the absolute phase shift of a weak probe passing through pumped media often measure the zeroth-order coefficient $a_0(T)$ in this expansion \cite{chen2007,lavorel2016}. Pump-probe studies in gases and solids have previously measured the time delay-dependent frequency shifts \cite{bartels2001,neradovskaia2022}, which corresponds to the linear coefficient $a_1(T)$. Four-wave mixing experiments in liquids have also measured the full amplitude and phase of the signal \cite{gallagher1998, hershberger2011}. Such a complete measurement of the emitted E-field phase gives access to both the absolute phase shift and other higher-order terms, especially chirp \cite{walz2022}, which is used as the main observable in our study involving gas-phase molecules.

\begin{figure}[t]
\centering
\includegraphics[width= 1.0 \textwidth]{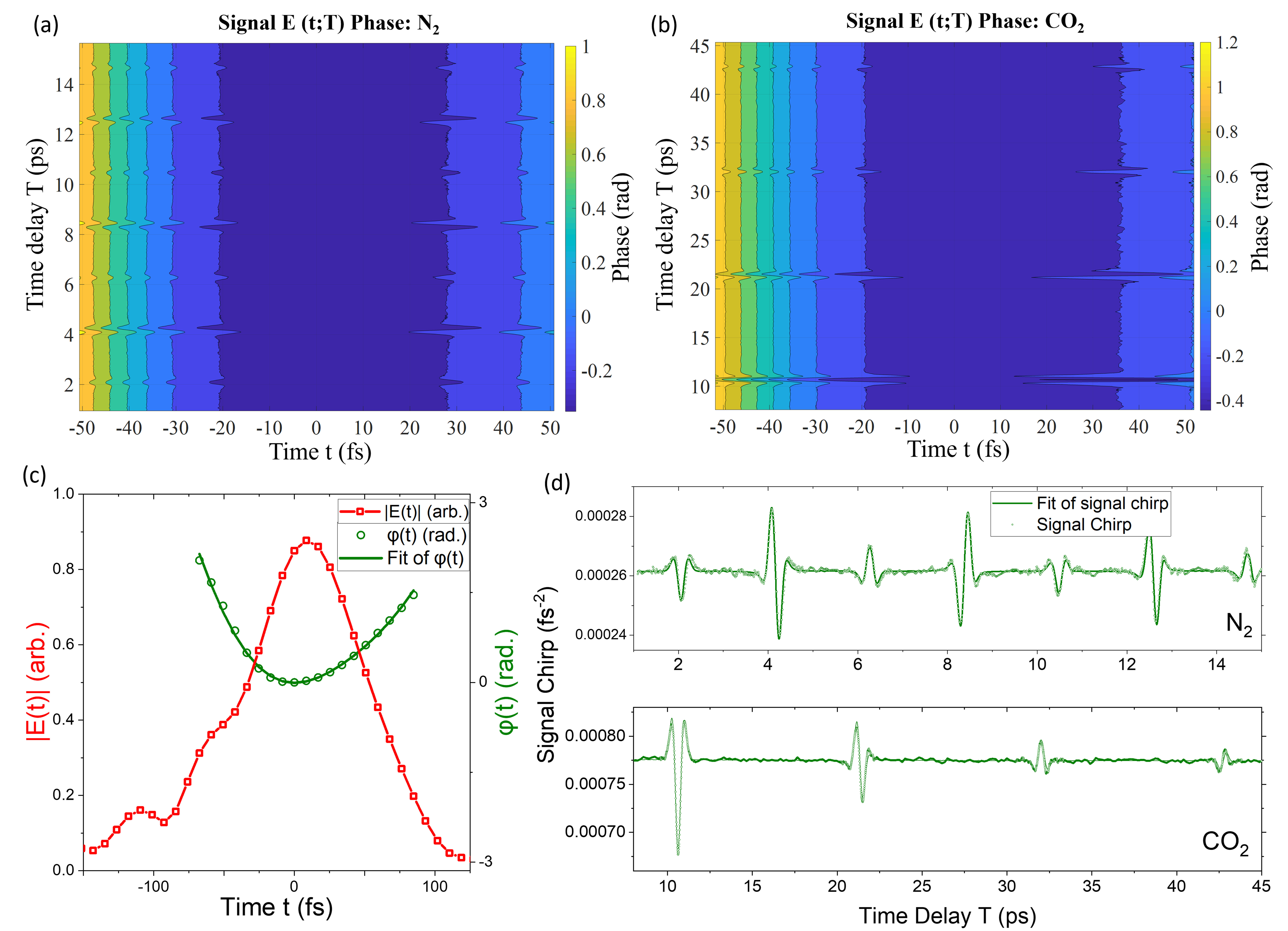}
\caption{(a) Temporal phase of the nonlinear signal E-field from pre-aligned N$_2$ molecules and (b) from pre-aligned CO$_2$ molecules, as a function of pulse time (t) and time delay (T). (c) A representative plot of E-field amplitude and phase along with a polynomial fit of the phase. While the actual temporal resolution of the measurement is 28 fs, the data is interpolated to aid visualization. (d) For each pump-probe time delay T, the measured E-field phase is fit with a polynomial in pulse time t. The second-order fit coefficient (chirp) is plotted as a function of T for N\textsubscript{2} and CO\textsubscript{2}. The chirp of the input probe pulses is $0.00017\, \mathrm{fs}^{-2}$.}
\label{fig2:fitting}
\end{figure}

\section{Results and Discussion}
Previous studies have shown that by measuring the photoionization or high-harmonic generation (HHG) yield from a molecular wavepacket as a function of alignment pump-probe time delay, the yield can be retrieved as a function of the relative angle between the pump pulse polarization and the symmetry-axis of the molecule \cite{ren2013,makhija2016,wang2017,makhija2019,sandor2018,sandor2019,lam2020}. This deconvolution method can improve angular resolution when working with molecular ensembles having low degree of alignment, as in our experiment where $\left\langle cos^{2}(\theta) \right\rangle$ $\sim$ 0.35. We perform such an analysis to retrieve the alignment angle dependence of the nonlinear signal E-field chirp from time delay-dependent measurements. We assume that the chirp of the emitted nonlinear signal is a function of the molecular alignment angle $\theta$. For linear molecules interacting with a one-color pulse, inversion symmetry implies $\theta \equiv \pi-\theta$, so we can expand the angle-dependent chirp $a_2(\theta)$ in Legendre polynomials as

\begin{equation}
    a_2(\theta) = \sum_l c_l \, P_l(\cos(\theta))
\end{equation}

with even values of $l$. On taking the expectation value of this equation with the pump-excited rotational wavepacket, the left-hand side becomes the experimentally measured chirp

\begin{equation}\label{eq:chirp}
    a_2(T) = \sum_j c_j \, \langle P_j \rangle (T)
\end{equation}

\begin{figure}[t]
\centering
\includegraphics[width= 1.0 \textwidth]{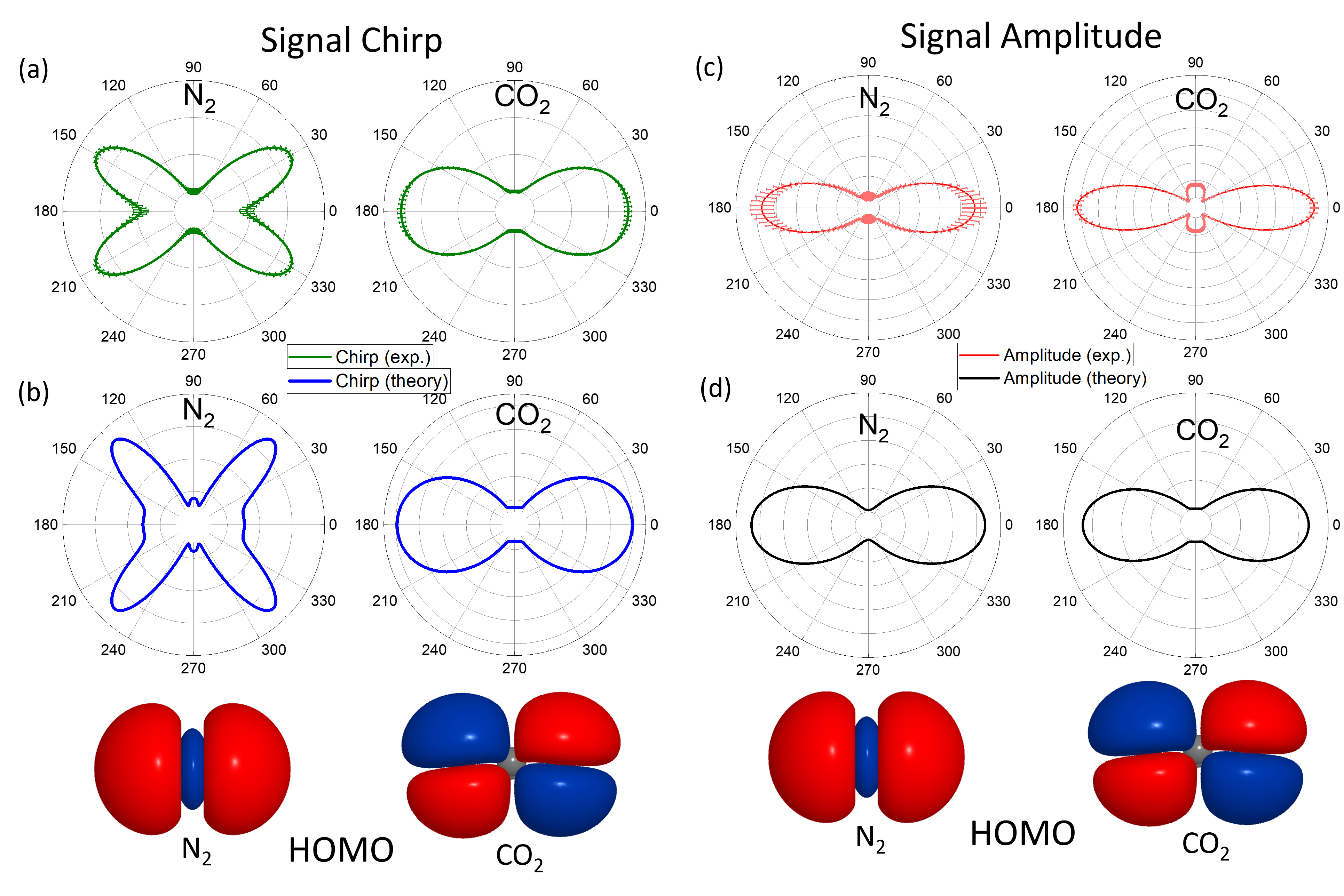}
\caption{(a) Alignment angle-dependent molecular-frame nonlinear signal E-field chirp retrieved from the experimental data, for N\textsubscript{2} and CO\textsubscript{2}. (b) Theoretical calculations of the alignment angle-dependent E-field chirp for N\textsubscript{2} and CO\textsubscript{2}. (c) alignment angle-dependent molecular-frame E-field amplitude (pulse time-integrated). (d) same as (c) from theoretical calculations. Highest occupied molecular orbitals (HOMO) for N\textsubscript{2} and CO\textsubscript{2} molecules, showing their distinct $\sigma$ and $\pi$ bonding character, respectively, are shown in the bottom panel}
\label{fig3:angular}
\end{figure}

Using a suitable set of pulse parameters for the alignment pump, we simulate the time evolution of the excited rotational wavepacket and calculate the expectation value of the Legendre polynomials on the right-hand side of equation \ref{eq:chirp} which can then be inverted to find the expansion coefficients $c_j$ (see reference \cite{makhija2016,wang2017} for more details). The rotational temperature of the gas is the same as its thermal temperature (295 K), and the pump pulse duration is measured to be 60 fs using a commercial Frequency Resolved Optical Gating (FROG) device. In the fitting procedure, the intensity of the pump pulse in the focal region was allowed to vary within reasonable bounds, from 5 to 40 TW/cm$^2$. To account for collisional dephasing of the excited rotational wavepacket, we also include a single-exponential decay parameter in the fitting procedure \cite{ramakrishna2006}. The measured signal E-field phase, as a function of pulse time (t) and pump-probe time delay (T), is shown in Fig. \ref{fig2:fitting} (a) for N$_2$ molecules and in Fig. \ref{fig2:fitting} (b) for CO$_2$ molecules, as contour plots. Figure \ref{fig2:fitting} (c) shows a representative plot of the pulse time (t) dependent amplitude and phase of the nonlinear signal for a fixed time delay (T). The phase is fit to a polynomial (equation \ref{eqn:phase}) using an amplitude weighted fit from which the chirp is extracted as a function of T. Figure \ref{fig2:fitting} (c) shows the extracted chirp as a function of T for both N$_2$ and CO$_2$ molecules. The alignment angle-dependent chirp in the molecular frame is retrieved by using a fitting and inversion procedure that provides coefficients $c_j$, as described above. Figure \ref{fig3:angular} (a) shows the molecular-frame chirp of the nonlinear optical signal for N$_2$ and CO$_2$ molecules. These experimental chirp plots show distinct angular dependence of the chirp for the two molecules which have different ground state electronic symmetries. The corresponding highest occupied molecular orbital (HOMO) for the two molecules are shown in the bottom panel of Fig. \ref{fig3:angular}. To investigate the origin of the angle-dependent chirp, the nonlinear electronic response of N\textsubscript{2} and CO\textsubscript{2} molecules was calculated using the DFWM pulse sequence used in the experiment. From the calculated signal E-fields, the phase was fit similar to the experimental data, to obtain the calculated molecular-frame angle-dependent chirp for a single molecule. The details of the calculation are provided in the next section. The chirp of the calculated signal field, as shown in Fig. \ref{fig3:angular} (b), shows good agreement with the experimentally determined angle-dependent chirp (Fig. \ref{fig3:angular} (a)). These calculations support the interpretation that the angle dependence of the signal chirp observed in our experiment is electronic in origin and that propagation effects of the weak nonlinear signal in the dense target medium are negligible. The low-intensity, non-resonant probe pulses used in this experiment interact perturbatively with the target molecules and signal predominantly originates from valence electrons. The differences between the molecular-frame chirp of the nonlinear signals from N\textsubscript{2} and CO\textsubscript{2} are thus sensitive to the differences in their valence electronic character. 

We perform a similar analysis to retrieve the alignment angle dependence of the (pulse time-integrated) amplitude of the emitted signal E-field. Figure \ref{fig3:angular} (c) shows that the amplitude of the measured signal is not sensitive to electronic character differences between the two molecules. The corresponding single-molecule theoretical calculations of the amplitude agrees well with the experimental data. This demonstrates that field-resolved nonlinear spectroscopy offers new observables such as the chirp that are sensitive to the electronic character and offer information beyond measurement of the intensity of the signal. Below, we briefly provide an explanation for the difference in sensitivity to electronic symmetries for the chirp and amplitude observables. 

In linear molecules, the magnitude of $\gamma^{(2)}_{zzzz}$ is generally larger than any other component of the second-hyperpolarizability tensor, and therefore (see Supplemental section for details), the amplitude and phase of the probed third-order nonlinear response in the frequency domain may be approximated using equation \ref{eq:chi3} as (frequency dependence is not shown)

\begin{equation}\label{eqn:chi3_mag_approx}
    |\chi^{(3)}_{zzzz}(\theta)| \approx |\gamma^{(2)}_{zzzz}| \cos^4(\theta)
\end{equation}

\begin{equation}\label{eqn:chi3_phase_approx}
    \varphi (\theta) \approx \varphi_{zzzz} + \frac 3 2 \Bigg|\frac{\gamma^{(2)}_{zzxx}}{\gamma^{(2)}_{zzzz}}\Bigg| \varphi_{zzxx} \sin^2(2\theta) + \Bigg|\frac{\gamma^{(2)}_{xxxx}}{\gamma^{(2)}_{zzzz}}\Bigg| \varphi_{xxxx} \sin^4(\theta)
\end{equation}

It is seen from equations \ref{eqn:chi3_mag_approx} and \ref{eqn:chi3_phase_approx} that the magnitude of the frequency-domain lab-frame nonlinear response, which is proportional to the amplitude of the measured nonlinear signal, contains only the predominant second hyperpolarizability ($\gamma^{(2)}$) tensor component. Whereas, the frequency dependent phase, which is required to obtain the time-dependent phase (and hence chirp), contains additional terms with multiple tensor components of $\gamma^{(2)}$. This provides a possible explanation for the sensitivity of the molecular-frame nonlinear signal chirp to the valence electronic character while the amplitude shows the same angular behavior for the two linear molecules. Although the amplitude of this nonlinear optical response is well understood, more work is needed to better understand the origins of the phase of these tensor components, and their relation to electronic symmetries.

\section{Calculation details}
Electronic structure models of CO$_2$ and N$_2$ molecules were constructed using coupled-cluster singles and doubles (CCSD) method, with the 6-31G(d,p) Gaussian basis set, and solved using the Dalton software~\cite{dalton2014}. These models were constructed by selecting "bright" states which have finite transition dipole moments to the ground state, and energetically low-lying states which have finite transition dipole to those bright states. All the bright states in these models have $B_u$ symmetry. More details are given in the Supplemental section.

We performed theoretical calculations of the emitted signal E-field using Lindblad equation simulations in the time-domain, solving

\begin{equation}\label{eq:lindblad}
\dot{\rho}(t) = -\frac{i}{\hbar}[H(t),\rho(t)] + \mathcal{L}_D \rho(t)
\end{equation}

with the Hamiltonian

\begin{equation}
H(\vec{r},t) = \Omega + \vec{\mu} \cdot (\vec{E}_1(\vec{r},t) + \vec{E}_2(\vec{r},t) + \vec{E}_3(\vec{r},t) )
\end{equation}

In these simulations, we use DFWM pulses $\vec{E}_1(\vec{r},t)$,  $\vec{E}_2(\vec{r},t)$,  $\vec{E}_3(\vec{r},t)$ with frequencies, durations, intensities, chirp, and polarizations that are the same as the experiment. The alignment pump is not included in the simulations as its only purpose was to align the molecules; alignment effects were captured in the simulations by rotating the DFWM pulses in the molecular frame.  Excited state energy levels $\Omega$ and transition dipole moments $\vec{\mu}$ are obtained from CCSD calculations. We included population relaxation and dephasing times of 1 ps in the Lindbladian $\mathcal{L}_D$, however, the results were insensitive to these values as the signal is non-zero only during the duration of pulse overlap, which is much shorter than the dephasing and relaxation times. The Lindblad equation was numerically solved using the Euler method with fixed time step of 0.1 fs, using the UTPS simulation package~\cite{utps2023}. 

The result of solving Eq.~\ref{eq:lindblad} is the time domain polarization $\vec{P}(\vec{r},t) = \mathrm{Tr}[\vec{\mu} \rho(\vec{r},t)]$. To extract the third-order nonlinear signal electric field, we impose phase matching conditions by selecting only wavevectors parallel to the signal propagation direction

\begin{equation}
\vec{P}_{\mathrm{sig}}(t) = \int \!\! d^3r \,\, e^{-i\vec{k}\cdot\vec{r}} \vec{P}(\vec{r},t)   
\end{equation}

with $\vec{k}=\vec{k_1}-\vec{k_2}+\vec{k_3}$ being the signal wavevector corresponding to the phase matching conditions of Eq.~\ref{eq:esignal}. These calculations were repeated for 100 alignment angles of the molecules between $0^\circ$ and $180^\circ$. Signal amplitude and chirp from these simulations were fit using the same methodology described above for the experimental data.

\section{Conclusion}
Field-resolved ultrafast spectroscopy is emerging as a sensitive approach to measure ultrafast dynamics on femtosecond and sub-femtosecond time scales in various systems. While recent studies have used field-resolved ultrafast measurement in solids \cite{sederberg2020} and liquids \cite{srivastava2023}, to our knowledge, no previous work has demonstrated field-resolved perturbative nonlinear spectroscopy in laser excited pre-aligned molecules in the gas phase. Multiple previous studies have successfully retrieved the angular structure of valence electronic orbitals in molecules using either photoionization spectroscopy \cite{litvinyuk2003,lin2006,pavivcic2007,thomann2008,marceau2017}, or by measuring emission of HHG light \cite{itatani2004,vozzi2005,vozzi2006,haessler2010,vozzi2011}. In this work, we have shown that the angle dependence of the measured third-order nonlinear electronic response in molecules can act as a probe of their valence electronic symmetry. By comparing the angle dependence of the measured E-field chirp and amplitude, we have found that the phase of the emitted nonlinear E-field can be more sensitive than the amplitude of the emitted signal, to the electronic symmetry of molecules. Further, we have demonstrated sensitivity of the nonlinear E-field chirp to the electronic character in molecules with poor degree of alignment ($\left\langle cos^{2}(\theta) \right\rangle$ $\sim$ 0.35) at room temperature, in a perturbative interaction not involving ionization, which has not been previously possible. Our experimental data are well-supported by theoretical calculations on the single molecule nonlinear response. A more detailed mechanistic understanding of the heightened sensitivity of E-field chirp to electronic nonlinearities is still needed.

The experiment presented here is a first step towards applying field-resolved measurements to study electronically excited states in atoms, molecules and solids, which opens up the possibility to disentangle complex quantum dynamics in real-time with unprecedented temporal resolution. Additionally, the sensitivity of the E-field phase to electronic symmetry, as demonstrated in the present work, provides a tool to study the transient changes in symmetry of electronic states in molecules as they evolve on excited potential energy surfaces. The ability to measure ultraweak fields with zeptojoule energies without delay scanning makes spectral interferometry \cite{fittinghoff1996} a suitable candidate for E-field metrology in experiments involving a low-intensity pump, such as a pulse from a HHG source, although direct field sampling has recently been demonstrated at the sub-femtojoule level \cite{srivastava2023}. In the future, measurement of field-resolved nonlinear optical signals from electronic states excited by a HHG source could offer new observables not previously accessible for the study of ultrafast dynamics.

\begin{backmatter}

\bmsection{Acknowledgments}
This material is based upon work supported by the National Science Foundation under Grant No. 2208061. Work at the Molecular Foundry was supported by the Office of Science, Office of Basic Energy Sciences, of the U.S. Department of Energy under Contract No. DE-AC02-05CH11231. This research used resources of the National Energy Research Scientific Computing Center, a DOE Office of Science User Facility supported by the Office of Science of the U.S. Department of Energy under Contract No. DE-AC02-05CH11231.

\bmsection{Disclosures}
The authors declare no conflicts of interest.

\bmsection{Data availability} Data underlying the results presented in this paper are not publicly available at this time but may be obtained from the authors upon reasonable request.

\bmsection{Supplemental document} See Supplement 1 for supporting content.

\end{backmatter}

\bibliography{references}

\end{document}


\maketitle

\section{Error Analysis of the Angle-dependent Chirp}
We make 25 independent measurements of the E-field amplitude and phase as a function of the pump-probe time delay (T). For each time delay, the E-field phase is fit with a polynomial in time (t), weighted by the normalized E-field intensity ($|E(t)|^2$). Each measurement then gives a T-dependent chirp ($a_2(T)$), which is averaged over all measurements to get T-dependent mean and standard error values for the chirp. A similar procedure is used for the E-field amplitude integrated over the pulse time (t).

\begin{figure}[h]
\centering
\includegraphics[width=\linewidth]{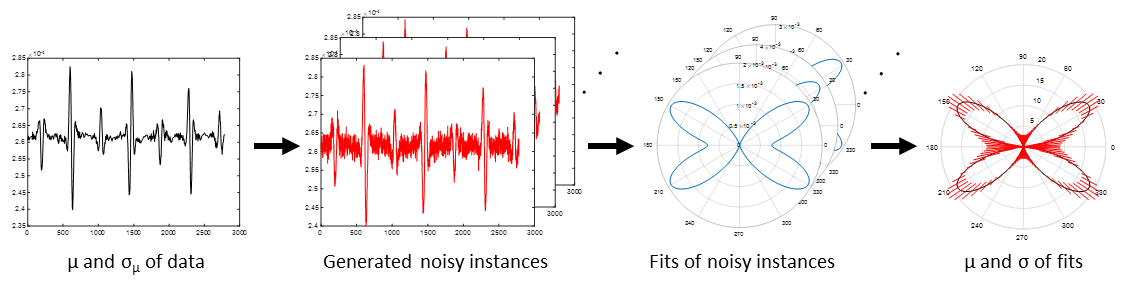}
\caption{Representation of the Monte Carlo uncertainty analysis: For each delay step, the mean and standard error of the experimental data (chirp or integrated amplitude) is used to generate random noisy instances of the T-dependent data. These instances are fit individually, and the retrieved angle-dependence from all these fits are used to generate the mean and uncertainty of the angle-dependence.}
\label{fig:monte_carlo}
\end{figure}

To extract the alignment angle-dependent chirp, we use the fitting procedure outlined in the main article. To estimate the uncertainty in the chirp we use a Monte Carlo approach. Starting with the mean T-dependent chirp, we generate noisy instances of the T-dependent chirp. The noise is assumed to be normally distributed, with a standard deviation equal to the standard error of the experimentally measured mean chirp. The fitting procedure is repeated for 1000 such noisy instances, from which the mean and standard deviation of the angle-dependent chirp are calculated and plotted. Figure \ref{fig:monte_carlo} graphically shows the error analysis procedure.


\section{Details of theoretical calculations}
The CO$_2$ model consists of the ground state together with the excited state singlet molecular orbitals of following symmetries: $B_{1g}$ (13.38 eV), $B_{2g}$ (10.69 eV), $B_{3g}$ (10.69 eV), $B_{2u}$ (15.33 eV), $B_{3u}$ (15.33 eV), two $B_{1u}$ orbitals (9.26 and 13.14 eV), and two $A_u$ orbitals (9.18 and 9.26 eV). The N$_2$ model consists of the ground state together with the excited state singlets: $B_{2g}$ (9.74 eV), $B_{3g}$ (9.74 eV), $B_{1u}$ (17.07eV), $B_{2u}$ (14.01 eV), $B_{3u}$ (14.01 eV). Since we do not expect significant ionization at the laser intensities used in the experiment, higher-lying ionic states were not included in the model.


\section{Approximating third-order nonlinear response amplitude and phase}
We start with the relevant lab-frame frequency-domain nonlinear response tensor component, expressed in terms of the molecular-frame second-hyperpolarizabilities $\gamma^{(2)}_{ijkl}$ and the relative angle between the laser polarization along $\hat{z}$ and the molecules' symmetry-axis:

\begin{equation}\label{eq:chi3}
    \chi^{(3)}_{zzzz}(\theta) = \gamma^{(2)}_{zzzz} \cos^4(\theta) + \frac 3 2 \gamma^{(2)}_{zzxx} \sin^2(2\theta) + \gamma^{(2)}_{xxxx} \sin^4(\theta)
\end{equation}

Writing all tensor components as complex numbers,

\begin{equation}
    |\chi^{(3)}_{zzzz}| e^{i\varphi} = |\gamma^{(2)}_{zzzz}| e^{i\varphi_{zzzz}} \cos^4(\theta) + \frac 3 2 |\gamma^{(2)}_{zzxx}| e^{i\varphi_{zzxx}} \sin^2(2\theta) + |\gamma^{(2)}_{xxxx}| e^{i\varphi_{xxxx}} \sin^4(\theta)
\end{equation}

which for small phase angles ($e^{i\varphi} \approx 1 + i\varphi$) becomes

\begin{equation}
    |\chi^{(3)}_{zzzz}| = |\gamma^{(2)}_{zzzz}| \cos^4(\theta) + \frac 3 2 |\gamma^{(2)}_{zzxx}| \sin^2(2\theta) + |\gamma^{(2)}_{xxxx}| \sin^4(\theta)
\end{equation}

\begin{equation}
    |\chi^{(3)}_{zzzz}| \varphi = |\gamma^{(2)}_{zzzz}| \varphi_{zzzz} \cos^4(\theta) + \frac 3 2 |\gamma^{(2)}_{zzxx}| \varphi_{zzxx} \sin^2(2\theta) + |\gamma^{(2)}_{xxxx}| \varphi_{xxxx} \sin^4(\theta)
\end{equation}

When $|\gamma^{(2)}_{zzzz}|$ is the dominant second-hyperpolarizability component, these equations can be simplified as

\begin{equation}
    |\chi^{(3)}_{zzzz}| \approx |\gamma^{(2)}_{zzzz}| \cos^4(\theta)
\end{equation}

\begin{equation}
    \varphi (\theta) \approx \varphi_{zzzz} + \frac 3 2 \Bigg|\frac{\gamma^{(2)}_{zzxx}}{\gamma^{(2)}_{zzzz}}\Bigg| \varphi_{zzxx} \sin^2(2\theta) + \Bigg|\frac{\gamma^{(2)}_{xxxx}}{\gamma^{(2)}_{zzzz}}\Bigg| \varphi_{xxxx} \sin^4(\theta)
\end{equation}

This shows that the measured spectral phase (and hence temporal phase) is related to multiple components of the molecular frame hyperpolarizability tensor whereas the amplitude depends mainly on the dominant component. This provides an explanation for the sensitivity of chirp to electronic symmetries and the lack of such sensitivity for the amplitude.










